\documentclass[12pt]{article}
\usepackage{epsf}
\newcommand{\bff}[1]{\mbox{\boldmath ${#1}$}}
\newcommand{\eps}{\epsilon}

\begin{document}

\begin{titlepage}
\begin{flushright}
 CERN-TH/99-57\\
 \mbox{DTP/99/26\hspace*{0.3cm}}\\
 hep-ph/9903260\\ 
 \mbox{March 1999\hspace*{0.5cm}}\\
\end{flushright}
\vskip 1.8cm
\begin{center}
\boldmath
{\Large\bf Top quark production near threshold\\[0.2cm] 
and the top quark mass}
\unboldmath 
\vskip 1.2cm
{\sc M. Beneke}
\vskip .3cm
{\it Theory Division, CERN, CH-1211 Geneva 23, Switzerland}
\vskip 0.7cm
{\sc A. Signer}
\vskip .3cm
{\it Department of Physics, University of Durham, \\ 
Durham DH1 3LE, England}
\vskip .7cm
{\sc V.A. Smirnov}
\vskip .3cm
{\em Nuclear Physics Institute, Moscow State University, \\
119889 Moscow, Russia}
\vskip 1.5cm

\end{center}

\begin{abstract}
\noindent 
We consider top-anti-top production near threshold in $e^+ e^-$ 
collisions, resumming Coulomb-enhanced corrections at 
next-to-next-to-leading order (NNLO). We also sum potentially large 
logarithms of the small top quark velocity 
at the next-to-leading logarithmic level using the renormalization 
group. The NNLO correction to the cross section is large, and it 
leads to a significant modification of the peak position and 
normalization. We demonstrate that an accurate top quark mass 
determination is feasible if one abandons the conventional pole 
mass scheme and if one uses a subtracted potential and the corresponding 
mass definition. Significant uncertainties in the normalization 
of the $t\bar{t}$ cross section, however, remain. 
\end{abstract}

\vfill

\end{titlepage}


{\bf 1. Introduction.} 
The top quark mass is now known to be around $175\,$GeV with an 
accuracy of $5\,$GeV from the direct measurement at the Fermilab 
Tevatron Collider \cite{topmass}. Accurate mass 
determinations (with errors below 
1-2$\,$GeV) are difficult at hadron colliders. Despite the fact that 
orders of magnitudes more top quarks will be produced at the CERN 
Large Hadron Collider, a precision measurement is reserved for a 
future lepton collider. In this case the method of choice relies on 
scanning the top quark pair production threshold. From an experimental 
point of view, an error in the $100\,$MeV range is conceivable 
\cite{tesla}; 
the limiting factor is the accuracy to which the cross section can 
be predicted theoretically as a function of the top quark mass.

The literature on top quark physics near threshold in $e^+ e^-$ 
collisions is substantial \cite{FK87,toplit}. 
Perturbative calculations in the 
threshold region require that either the toponium Rydberg energy 
scale $m_t \alpha_s^2\gg \Lambda_{\rm QCD}$ or that the top quark 
decay width $\Gamma_t\gg \Lambda_{\rm QCD}$. Both conditions are 
satisfied and since $\Gamma_t\sim m_t\alpha_s^2\sim 1.5\,$GeV, 
narrow toponium resonances do not exist \cite{B86}. In the kinematic 
region of interest, the top quarks are slow, with typical velocities 
$v\sim 1/10$. As a consequence methods familiar from 
non-relativistic bound state calculations in QED can be used 
to compute the 
$t\bar{t}$ cross section. In particular, the dominant interaction 
between the $t$ and $\bar{t}$ can be described by the colour-singlet 
Coulomb potential, which has to be treated non-perturbatively 
\cite{FK87}. Corrections are computed in the background of this strong 
Coulomb interaction.

Recently, the 2-loop QCD correction to the Coulomb potential 
\cite{Sch98} (correcting an earlier result \cite{Pet97}) and the 
2-loop relativistic correction to the $t\bar{t}$ 
vector coupling to the initial virtual photon or $Z$ boson 
\cite{BSS98,CM98} have been computed. 
With these two inputs at hand, we can compute 
the next-to-next-to-leading order (NNLO) QCD correction to the 
top quark pair production cross section. (The precise meaning of 
`NNLO' in the present context is given below.) In the 
following, we first compute the NNLO QCD correction in the 
conventional pole mass scheme, i.e. the threshold cross section is 
expressed in terms of the top quark pole mass. Comparable 
calculations, with some technical and implementational differences, 
have already been completed by several groups \cite{HT98,MY98,Yak98}. 
We find that the NNLO correction is substantial 
and leads to an uncomfortably 
large shift in the top quark mass, raising questions as to the 
stability of the theoretical prediction. In \cite{Ben98}, one of us 
suggested that such shifts could occur as a consequence of on-shell 
mass renormalization, which is particularly (and in the case at hand, 
artificially) sensitive to non-perturbative, long-distance effects.
This sensitivity to non-perturbative effects can be removed 
by using a different mass renormalization scheme (called the 
potential subtraction (PS) scheme in \cite{Ben98}) with the additional 
benefit that the new mass definition can also be related more 
reliably to short-distance masses (such as the $\overline{\rm MS}$ 
mass), which are ultimately of more interest for high-energy 
processes and Yukawa coupling unification relations (should they 
exist). 

The main result of this letter is to demonstrate that this 
procedure works. We show that the mass shifts become small in the 
PS scheme, and that the PS mass can also be accurately related to the 
$\overline{\rm MS}$ mass. However, the normalization uncertainty 
of the cross section remains large, and is larger than could have 
been anticipated from the earlier 
next-to-leading order (NLO) calculations. The second improvement 
which we suggest in this letter is to use renormalization group 
equations to 
sum large logarithms, $\ln v$ and $\ln v^2$ to all orders 
in perturbation theory. The consideration 
of logarithms is important to identify correctly the scales of the various 
subprocesses, but not restricted to the scale of the 
QCD coupling only. In the present 
work, we complete this program (almost trivially) to the 
next-to-leading logarithmic (NLL) order. The step to NNLL is 
substantially more complicated and we will discuss it, together 
with the details of the present calculation, in a future 
publication \cite{inprep}.

Before proceeding, let us emphasize that `NLO', `NNLO' etc. 
do not refer to a conventional loop expansion in $\alpha_s$, 
because the Coulomb interaction cannot be treated perturbatively 
in the threshold region. To be precise, a LO, NLO, ... approximation 
to the normalized cross section takes into account all terms of 
order
\begin{equation}
\label{syst}
R\equiv\sigma_{t\bar{t}}/\sigma_{\mu^+\mu^-} = 
v \,\sum_k \left(\frac{\alpha_s}{v}\right)^k \cdot 
\left\{1\,\mbox{(LO)};\,\alpha_s,v\,\mbox{(NLO)};\,
\alpha_s^2,\alpha_s v,v^2\,\mbox{(NNLO)};\,\ldots\right\},
\end{equation}
where logarithms of $v$ are suppressed. The renormalization group 
improved treatment extends this to the summation of logarithms 
such that a LL, NLL, ... approximation accounts for all terms 
of order 
\begin{equation}
\label{systrge}
R = 
v \,\sum_k \left(\frac{\alpha_s}{v}\right)^k 
\sum_l \left(\alpha_s \ln v\right)^l \cdot 
\left\{1\,\mbox{(LL)};\,\alpha_s,v\,\mbox{(NLL)};\,
\alpha_s^2,\alpha_s v,v^2\,\mbox{(NNLL)};\,\ldots\right\}.
\end{equation}
The result discussed here is complete at `NNLO' and `NLL'. 
Furthermore, near the would-be toponium poles another resummation 
is necessary, which we discuss below.

We also emphasize that the NNLO QCD correction is defined as above 
in the limit $\Gamma_t\ll m_t \alpha_s^2$. When $\Gamma_t \sim 
m_t\alpha_s^2$, as we expect, further corrections arise from 
so-called non-factorizable contributions \cite{nonfact}, which have not 
been calculated so far with the Coulomb interaction treated 
non-perturbatively. Electroweak corrections also enter and the entire 
concept of a $t\bar{t}$ cross section has to be revised, since 
single resonant contributions are expected to contribute at NNLO 
(in the above power counting scheme with 
$\Gamma_t \sim m_t\alpha_s^2$) to the $WWb\bar{b}$ final state. 
These are interesting issues to be studied, but we expect them to 
bear little on the issue of an accurate top mass determination which we 
address in this letter. Hence, we will keep the top quark width 
only in the form of an imaginary mass term $i\Gamma_t\psi^\dagger \psi$ 
for the non-relativistic top quarks; this amounts to evaluating 
the $t\bar{t}$ Green function at energy 
$E+i\Gamma_t$ as familiar from LO and NLO 
calculations \cite{FK87,toplit}. As a final (trivial) simplification, 
we restrict our attention to $t\bar{t}$ pairs produced through 
the vector coupling to a virtual photon. The vector coupling to 
a $Z$ boson can be accommodated by a trivial replacement of the 
electric charge. The axial vector coupling is suppressed by a factor 
of $v^2$ near threshold; no QCD corrections to it are needed at 
NNLO.

\vspace*{0.2cm}


{\bf 2. Derivation of the cross section.} 
With the treatment of the top quark width as specified above, and 
neglecting the axial-vector coupling, the 
$t\bar{t}$ production cross section is obtained from the 
correlation function 
\begin{equation}
\label{pi}
\Pi_{\mu\nu}(q^2) = (q_\mu q_\nu-q^2 g_{\mu\nu})\,\Pi(q^2) = 
i\int d^4 x\,e^{i q\cdot x}\,\langle 0|T(j_\mu(x) j_\nu(0))|0\rangle,
\end{equation}
where $j^\mu(x) = [\bar{t}\gamma^\mu t](x)$ is the top quark vector 
current and $s=q^2$ the centre-of-mass energy squared. Defining 
the usual $R$-ratio $R=\sigma_{t\bar{t}}/\sigma_0$ ($\sigma_0=
4\pi \alpha_{em}^2/(3 s)$, where $\alpha_{em}$ is the electromagnetic 
coupling at the scale $2 m_t$), the relation is 
\begin{equation}
R = \frac{4\pi e_t^2}{s}\,\mbox{Im}\,\Pi^{ii}(s+i\epsilon),
\label{rr1}
\end{equation}
where $e_t=2/3$ is the top quark electric charge in units of the positron 
charge. Only the spatial components of the currents contribute up to NNLO. 
In the following, $m_t$ refers to the top quark pole mass.

According to (\ref{syst}), at NNLO, we have to extract, to all orders 
in $\alpha_s$, the first three terms of the expansion of any Feynman 
diagram in $v=((\sqrt{s}-2 m_t)/m_t)^{1/2}$. The expansion in $v$ 
is constructed \cite{BS98} by dividing the loop integral into terms 
related to hard ($l_0\sim m_t$, $\bff{l}\sim m_t$ -- referring to a frame 
where $\bff{q}=0$), soft ($l_0\sim m_t v$, $\bff{l}\sim m_t v$), 
potential ($l_0\sim m_t v^2$, $\bff{l}\sim m_t v$) and ultrasoft 
($l_0\sim m_t v^2$, $\bff{l}\sim m_t v^2$) momentum. This split-up requires 
a regularization to deal with divergent integrals that appear in 
intermediate expressions and we use dimensional regularization with 
$\overline{\rm MS}$ subtractions. This procedure has already been used 
to obtain the cross section at threshold up to order $\alpha_s^2$. 
By expanding the all-order result presented below up to order $\alpha_s^2$, 
we reproduced the result of \cite{CM98}, which has been used as a common 
input to the previous \cite{HT98,MY98,Yak98} NNLO $t\bar{t}$ cross 
section calculations.

We begin with integrating out the hard modes, commonly termed 
`relativistic corrections'. The effective $\gamma^*t\bar{t}$ coupling 
seen by the non-relativistic quarks (described by two-spinor fields 
$\psi$ for $t$ and $\chi$ for $\bar{t}\,$) after integrating out the 
hard modes is given by 
\begin{equation}
\label{current}
\bar{t}\gamma^i t = c_1\,\psi^\dagger \sigma^i\chi - 
\frac{c_2}{6 m_t^2}\,\psi^\dagger \sigma^i (i\bff{D})^2\chi + 
\ldots,
\end{equation}
where the ellipsis refers to terms not needed for $\Pi(q^2)$ and at 
NNLO. At NNLO, we can use $c_2=1$, while $c_1$ is needed at 
order $\alpha_s^2$. The expression for $c_1$ is
\begin{equation}
\label{cc1}
c_1(\mu) = 1+ \left[(c_1^{(1)}+\delta_1)\,\frac{\alpha_s(\mu_h)}
{\alpha_s(\mu)}-\delta_1\right] \frac{\alpha_s(\mu)}{4\pi}
+c_1^{(2)} \frac{\alpha_s^2}{(4\pi)^2} + \ldots 
\end{equation}
where $\delta_1$, related to the 2-loop anomalous dimension of the 
non-relativistic current $\psi^\dagger \sigma^i\chi$, is given by 
$\delta_1 = -560\pi^2/(27 b_0)$ ($b_0=11-2 n_f/3$, $n_f=5$) 
and $c_1$ at the scale $\mu_h$, at which QCD is matched onto a 
non-relativistic effective theory, is given by
\begin{eqnarray}
\label{cc2}
c_1(\mu_h) &=& 1- \frac{8\alpha_s(\mu_h)}{3\pi} + \Bigg[
\left\{\frac{2}{3}\,b_0+\frac{35\pi^2}{27}\right\} 
\ln\frac{m_t^2}{\mu_h^2} 
\nonumber\\
&&-\,\frac{89}{54}-\frac{511\pi^2}{324}-\frac{14\pi^2}{9} 
\ln 2-\frac{125\zeta(3)}{9}+\frac{11}{27}\,n_f \Bigg] 
\frac{\alpha_s(\mu_h)^2}{\pi^2} + \ldots.
\end{eqnarray}
The first order result is well known \cite{BGKK75}; the second order 
contribution is from \cite{BSS98,CM98}. Eq.~(\ref{cc1}) follows from 
solving the renormalization group equation for $c_1$ with the 
2-loop anomalous dimension. When evaluated at a scale $\mu$ of order 
$m v$, this expression sums next-to-leading logarithms of the 
form $\alpha_s (\alpha_s \ln v)^l$ to all orders. This is in fact the 
only source of next-to-leading logarithms in the problem 
(there are no leading logarithms), and (\ref{cc1}) is sufficient to 
obtain a NLL approximation. There is an ambiguity in the scale of 
$\alpha_s^2$ in (\ref{cc1}). In our numerics, we actually choose 
the expression for the NNLL-improved coefficient function, setting 
the (unknown) 3-loop anomalous dimension of the current to zero.

The correlation function (\ref{pi}) is now expressed in terms of 
the effective current (\ref{current}). (This leaves out a hard correction 
from the region $x\sim 1/m_t$ in the integral (\ref{pi}). However, 
in this region the top quarks are far off-shell and no contribution 
to the imaginary part of $\Pi$ is obtained.) The correlation functions 
of the non-relativistic current are then computed with the 
non-relativistic effective Lagrangian. It is a straightforward matter 
of counting powers of velocity (using the momentum scaling rules 
given above) to show that the following terms in the effective 
Lagrangian are sufficient at NNLO:
\begin{eqnarray}
\label{nrqcd}
{\cal L}_{\rm NRQCD} &=& 
\psi^\dagger \left(i D^0+\frac{\bff{D}^2}{2 m_t}+i\Gamma_t
\right)\psi + \frac{1}{8 m_t^3}\,\psi^\dagger\bff{D}^4\psi
-\frac{d_1\,g_s}{2 m_t}\,\psi^\dagger\bff{\sigma}\cdot \bff{B}\psi
\nonumber\\
&&\hspace*{-1.5cm}
+\,\frac{d_2\,g_s}{8 m_t^2}\,\psi^\dagger\left[D^i, E^i\right]\psi
+ \frac{d_3\,i g_s}{8 m_t^2}\,\psi^\dagger\sigma^{ij}\left[
D^i,E^j\right]\psi 
+\mbox{antiquark terms } + \,{\cal L}_{\rm light}.
\end{eqnarray}
Because we use dimensional regularization, some care is needed to define 
the algebra of Pauli matrices in $3-2\epsilon$ dimensions as well as 
anti-symmetric products. We use $\sigma^{ij}\equiv 
[\sigma^i,\sigma^j]/(2 i)$ (equal to $\epsilon^{ijk} \sigma^k$ in 
three dimensions) and $\bff{\sigma}\cdot\bff{B}$ must be 
interpreted as $-\sigma^{ij} G^{ij}/2$ in terms of the gluon field 
strength tensor. The last term in (\ref{nrqcd}), ${\cal L}_{\rm light}$, 
denotes the QCD Lagrangian of the massless fields, i.e. the QCD Lagrangian 
with the top quark part omitted. The coefficient functions $d_1$, $d_2$, 
$d_3$ can be set to 1 at NNLO. Their leading logarithmic renormalization 
would be required for a NNLL approximation. In that case, further operators, 
notably four-fermion operators (of heavy-heavy and heavy-light type) 
would have to be added to the effective Lagrangian. We discuss this 
extension in \cite{inprep}. The unconventional term 
involving the top quark width accounts for the fast top quark decay 
as discussed in the introduction. (We should emphasize again that the 
Lagrangian is not complete to NNLO as far as the treatment of the 
width is concerned. The term we added is the leading order term, but 
further terms exist which are suppressed by two powers of velocity.)

The loop integrals constructed with the non-relativistic Lagrangian still 
contain soft, potential and ultrasoft 
modes. Near threshold, where energies are of order $m_t v^2$, only 
potential top quarks and ultrasoft gluons (light quarks) can appear as 
external lines of a physical scattering amplitude. Hence, we integrate out 
soft gluons and quarks and potential gluons (light quarks) and 
construct the effective Lagrangian for the potential top quarks and 
ultrasoft gluons (light quarks). Because the modes that are integrated out 
have large energy but not large momentum compared to the modes we keep, 
the resulting Lagrangian contains instantaneous, but spatially non-local 
interactions. In the simplest case, these reduce to what is commonly 
called the `heavy quark potential'. We refer to this theory as 
potential non-relativistic QCD (PNRQCD), adapting the term PNRQED 
introduced in \cite{PS98} to the QCD case.

The derivation of the potentials in the framework of the threshold 
expansion \cite{BS98} will be presented elsewhere \cite{inprep}. The 
following result has been obtained by matching the on-shell $t\bar{t}$ 
scattering amplitude in NRQCD onto its PNRQCD counterpart. We verified 
that the potential is gauge-independent for a general covariant gauge 
and the Coulomb gauge. (To obtain a gauge invariant result one has to 
combine the contribution from the soft modes with the one from potential 
gluons.) The resulting momentum space potential, after carrying out a 
colour and spin projection on the components relevant to the 
calculation of $\Pi$, is
\begin{eqnarray}
\tilde{V}(\bff{p},\bff{q}) &=& -\frac{4\pi C_F\alpha_s}{\bff{q}^2} 
+\delta\tilde{V}(\bff{p},\bff{q}),
\\
\delta\tilde{V}(\bff{p},\bff{q}) &=&  
-\frac{4\pi C_F\alpha_s}{\bff{q}^2} \Bigg[
\left(a_1-b_0\ln\frac{\bff{q}^2}{\mu^2}\right) \frac{\alpha_s}{4\pi} 
\nonumber\\
&&\hspace*{-1.5cm}
+ \,\left(a_2-(2 a_1 b_0+b_1) \ln\frac{\bff{q}^2}{\mu^2}+
b_0^2 \ln^2\frac{\bff{q}^2}{\mu^2}\right) \frac{\alpha_s^2}{(4\pi)^2} 
\nonumber\\
&&\hspace*{-1.5cm}
+\,\frac{\pi\alpha_s|\bff{q}|}{4 m_t}\left(\frac{\bff{q}^2 e^{-\gamma_E}}
{\mu^2}\right)^{\!-\epsilon} 
\frac{\Gamma(1/2-\epsilon)^2\Gamma(1/2+\epsilon)}{\pi^{3/2}
\Gamma(1-2\epsilon)} 
\left(-\frac{C_F}{2} (1-2\epsilon)+C_A (1-\epsilon)\right)
\nonumber\\
&&\hspace*{-1.5cm}
+\,\frac{\bff{p}^2}{m_t^2}+\frac{\bff{q}^2}{m_t^2} 
\left\{\frac{d^2-7 d+10}{4 (d-1)}\,d_1^{\,2}- \frac{1}{4}(1+d_2)\right\}
\Bigg],
\label{delV}
\end{eqnarray}
where $d=4-2\eps$, $C_F=4/3$, $C_A=3$, $d_1=d_2=1$ in the present 
NNLO/NLL approximation, and $b_1=102-38 n_f/3$ the two-loop coefficient 
of the QCD $\beta$-function. (Always $\alpha_s=\alpha_s(\mu)$.) The 
loop corrections to the Coulomb potential are $a_1=(31 C_A/9-10 n_f/9)$ 
\cite{FB77} and \cite{Sch98}
\begin{eqnarray}
a_2&=&C_A^2\left[\frac{4343}{162}+\frac{22\zeta(3)}{3}+4\pi^2-\frac{\pi^4}{4}
\right] - \,C_A n_f \left[\frac{899}{81}+\frac{28\zeta(3)}{3}\right]
\nonumber\\
&&-\,C_F n_f\left[\frac{55}{6}-8\zeta(3)\right] + \frac{100 n_f^2}{81}.
\end{eqnarray}
The potential (\ref{delV}) differs from the potential used in 
\cite{HT98,MY98,Yak98}. Firstly, we need the potential in $d$ space-time 
dimensions, because the terms in the last two lines generate ultraviolet 
divergent integrals, which we regularize dimensionally. The divergences 
cause a factorization scale dependence which cancels with 
the factorization scale dependence in the coefficient function 
(\ref{cc1}) of the non-relativistic current. (The Coulomb potential 
does not generate divergences and we can use $a_{1,2}$ in four 
dimensions.) Secondly, our potential contains a $C_F^2\alpha_s^2/
(m_t\,|\bff{q}|)$ term, which is absent in \cite{HT98,MY98,Yak98}. 
Nevertheless, both potentials (in four dimensions) are in fact 
equivalent \cite{inprep}.

Having integrated out soft modes and potential gluons, the correlation 
functions of non-relativistic currents are computed with the 
Lagrangian
\begin{eqnarray}
\label{pnrqcd}
{\cal L}_{\rm PNRQCD}
 &=& 
\psi^\dagger \left(i \partial^0+\frac{\bff{\partial}^2}{2 m_t}+i\Gamma_t
\right)\psi +  
\chi^\dagger \left(i \partial^0-\frac{\bff{\partial}^2}{2 m_t}+i\Gamma_t
\right)\chi
\nonumber\\
&&\hspace*{-1.5cm}
+\, \int d^{d-1} r \left[\psi^\dagger\psi\right](r)
\left(-\frac{C_F\alpha_s}{r}\right) 
\left[\chi^\dagger\chi\right](0)
\nonumber\\
&&\hspace*{-1.5cm}
+ \, \frac{1}{8 m_t^3}\,\psi^\dagger\bff{\partial}^4\psi 
- \frac{1}{8 m_t^3}\,\chi^\dagger\bff{\partial}^4\chi 
+ \int d^{d-1} r \left[\psi^\dagger\psi\right](r)
\,\delta V(r)\, 
\left[\chi^\dagger\chi\right](0)
\end{eqnarray}
The terms in the last line are treated as perturbations. However, 
velocity power counting reproduces the well known fact that the 
leading order Coulomb potential in the second line is not suppressed 
compared to the free non-relativistic Lagrangian in the first line. 
Consequently perturbation theory in PNRQCD implies that instead of 
freely propagating, a $t\bar{t}$ pair propagates according to the 
Coulomb Green function, which satisfies
\begin{equation} 
\left(\frac{\bff{p}^2}{m_t}-\bar{E}\right)\tilde{G}_c(\bff{p},
\bff{p}';\bar{E}) + \int\!\frac{d^{d-1}\bff{k}}{(2\pi)^{d-1}}
\left(\frac{-4\pi C_F\alpha_s}{\bff{k}^2}\right) \tilde{G}_c(\bff{p}-\bff{k},
\bff{p}';\bar{E}) = (2\pi)^{d-1}\,\delta^{(d-1)}(\bff{p}-\bff{p}')
\end{equation}
with $\bar{E}=E+i\Gamma_t$ and $E=\sqrt{s}-2 m_t$.
Because we use dimensional regularization, all quantities are 
defined {\em a priori} in momentum space; the above equation 
defines the Coulomb Green function in $d$ dimensions. 
The PNRQCD Lagrangian (\ref{pnrqcd}) does not contain any gluon fields 
any more. This is so, because at NNLO the top quarks interact only 
through potentials. Counting powers of velocity for the leading 
ultrasoft interactions, we find that they are of NNNLO and higher 
order, i.e. beyond the accuracy of the present calculation. 

To complete the calculation, we compute with PNRQCD the correlation 
functions of the non-relativistic currents. For the power-suppressed 
term in (\ref{current}) the LO Lagrangian suffices. For the 
current $\chi^\dagger\sigma^i\psi$ we need the kinetic energy 
correction to first order, the potentials suppressed by one power 
of $\alpha_s$ or $v$ relative to the Coulomb potential to second order, 
and the potential suppressed by two powers of $\alpha_s$ or $v$ 
to first order. The explicit result for the cross section is lengthy 
and will be given in \cite{inprep}. We have checked this result 
by expanding it to order $\alpha_s^2$, confirming the result 
of \cite{CM98} in this limit. This gives us confidence that the 
factorization in dimensional regularization has been done correctly. 
The Coulomb Green function contains bound state poles at 
$E_n^{\rm LO}=-m_t (C_F\alpha_s)^2/(4 n^2)$ for positive integer $n$. 
The location and residues of the bound state poles are modified by 
QCD corrections. We computed the bound state energies to order 
$\alpha_s^4$ and residues to order $\alpha_s^5$ and find agreement 
with the results of \cite{PY97} and \cite{MY98II}, respectively. 

The calculation described so far sums correctly all terms at 
NNLO and NLL, as defined in (\ref{syst}) and (\ref{systrge}). 
However, the result contains terms of the schematic form 
\begin{equation}
\label{sing}
\left[\frac{\alpha_s E_n^{\rm LO}}{E_n^{\rm LO}-(E+i\Gamma_t)}\right]^k
\end{equation}
which become large in the vicinity of $E=E_n^{\rm LO}$, if $\Gamma_t$ is 
not much larger than $E_n^{\rm LO}$. 
For top quarks $\Gamma_t\sim E_1^{\rm LO}$ 
and it is necessary to sum singular terms of the form (\ref{sing}) 
to all orders. This is easily done by adding the expression 
with the exact bound state energy denominator at NNLO and 
by subtracting the same expression but expanded and truncated at NNLO.
That is, we add:
\begin{eqnarray}
\label{residues}
&&\frac{F_n^{\rm LO} \left(1+f_1 \alpha_s+f_2\alpha_s^2\right)}{
E_n^{\rm LO} \left(1+e_1 \alpha_s+e_2\alpha_s^2\right)-\bar{E}}
-\Bigg\{\frac{F_n^{\rm LO}}{E_n^{\rm LO}-\bar{E}} 
+ \alpha_s\left[-\frac{F_n^{\rm LO}E_n^{\rm LO} e_1}{(E_n^{\rm LO}-\bar{E})^2}
+\frac{F_n^{\rm LO}f_1}{E_n^{\rm LO}-\bar{E}}\right]
\nonumber\\
&&
+\, \alpha_s^2 \left[\frac{F_n^{\rm LO}(E_n^{\rm LO})^2 e_1^2}
{(E_n^{\rm LO}-\bar{E})^3}
-\frac{F_n^{\rm LO}E_n^{\rm LO} (e_2+e_1 f_1)}{(E_n^{\rm LO}-\bar{E})^2}
+\frac{F_n^{\rm LO}f_2}{E_n^{\rm LO}-\bar{E}}\right]
\Bigg\}.
\end{eqnarray}
This procedure 
(also discussed in \cite{MY98,PP98}) is essentially equivalent (near the 
bound state poles) to solving the Schr\"odinger equation exactly 
with the potential $\delta V(r)$, rather than treating it perturbatively 
as we have done so far. In practice, we implement this resummation 
for the first two bound state poles. The correction for the remaining 
ones is tiny, because the residues decrease as $1/n^3$. It is worth 
noting that even after this resummation, one would not be able to 
compute the $t\bar{t}$ cross section in the threshold region, if the 
width of the top quark were smaller than about two times the error 
that remains in the location of the 1S bound state pole at NNLO. In the 
conventional pole mass scheme this requires $\Gamma_t$ to be larger 
than roughly $1\,$GeV, a constraint which is satisfied but not by a 
large margin.

\vspace*{0.2cm}

\begin{figure}[p]
   \vspace{-5cm}
   \epsfysize=18cm
   \epsfxsize=12cm
   \centerline{\epsffile{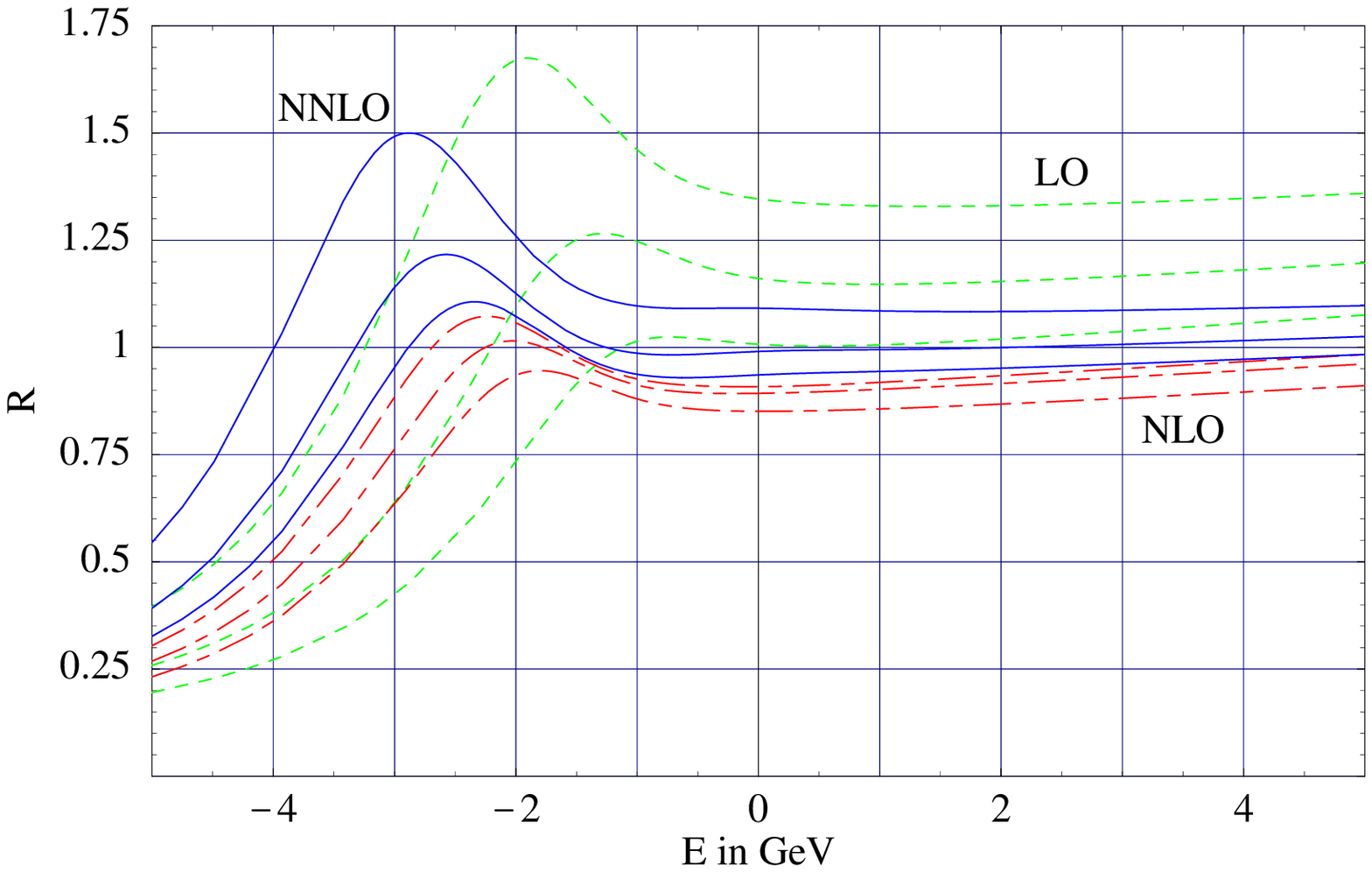}}
   \vspace*{-8.7cm}
   \hspace*{-0.1cm}
   \epsfysize=18cm
   \epsfxsize=12cm
   \centerline{\epsffile{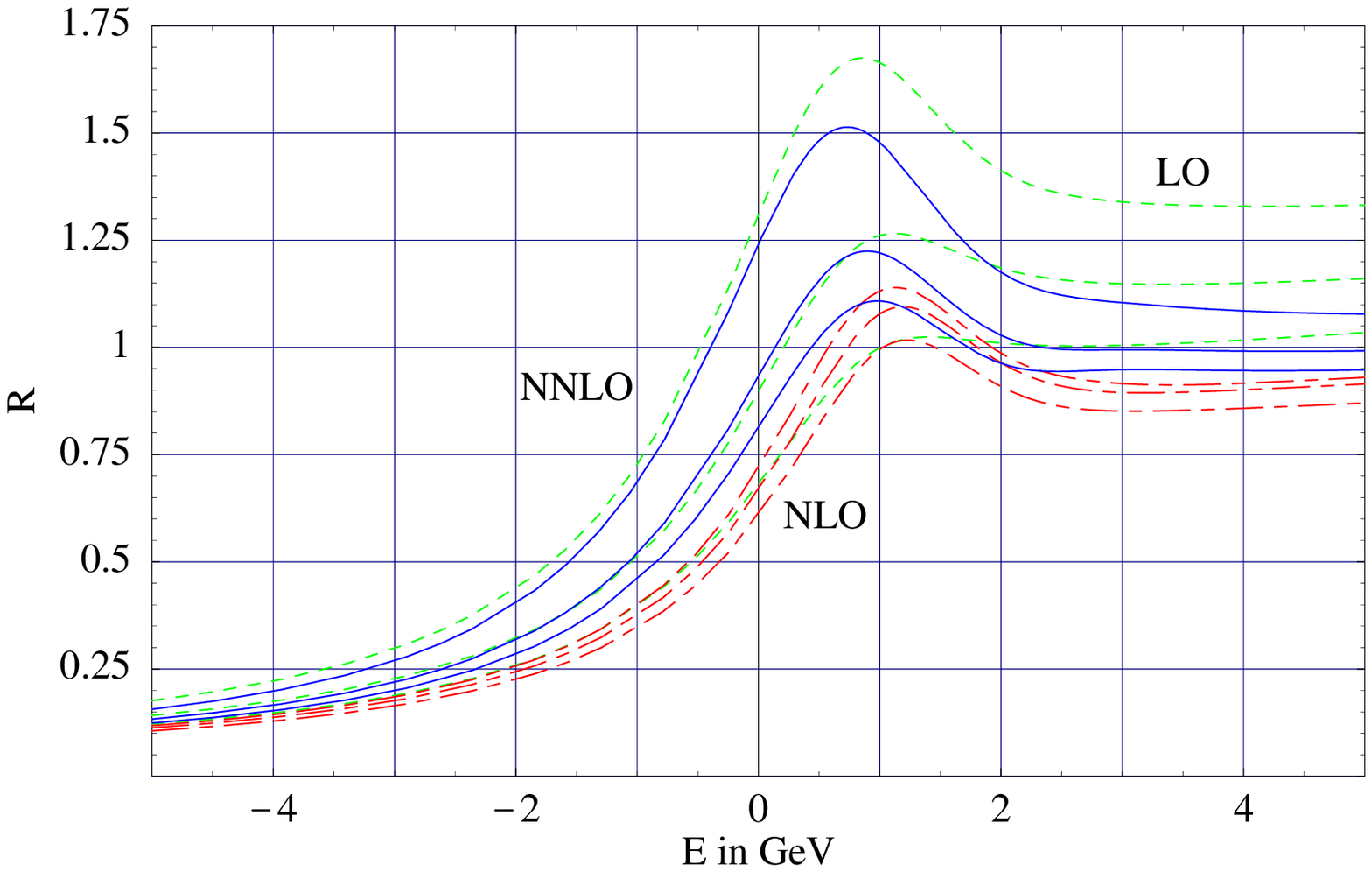}}
   \vspace*{-3.7cm}
\caption[dummy]{\label{fig1}\small 
(a) [upper panel]: The normalized $\bar{t}t$ cross section (virtual 
photon contribution only) in LO (short-dashed), NLO (short-long-dashed) 
and NNLO (solid) as function of $E=\sqrt{s}-2 m_t$ (pole mass scheme). 
Parameters: $m_t=\mu_h=175\,$GeV, $\Gamma_t=1.40\,$GeV, 
$\alpha_s(m_Z)=0.118$. The three curves for each case refer to 
$\mu=\left\{15 (\mbox{upper}); 30 (\mbox{central}); 
60 (\mbox{lower})\right\}\,$GeV. (b) [lower panel]: 
As in (a), but in the PS mass scheme with $\mu_f=20\,$GeV. Hence 
$E=\sqrt{s}-2 m_{t,\rm PS}(20\,\mbox{GeV})$. Other parameters as 
above with $m_t\to m_{t,\rm PS}(20\,\mbox{GeV})$.}
\end{figure}

{\bf 3. Top quark mass definitions, the PS scheme.}
The top quark cross section at LO, NLO and NNLO (including the 
summation of logarithms at NLL) is shown in Fig.~\ref{fig1}a. (To be 
precise, the NLO curves include the second iterations of the 
NLO potentials.) The NNLO correction is seen to modify the line 
shape at the level of 20\%. It also shifts the position of the 
peak by approximately $600\,$MeV. This conclusion is in qualitative 
agreement with the results of \cite{HT98,MY98,Yak98}. We should note, 
however, that our result is implemented in a different way: for 
instance, we do not keep the short-distance coefficient $c_1(\mu)^2$ 
as an overall factor, but multiply it out, keeping all terms to NNLO. 
Furthermore, we prefer to choose a different renormalization scale 
(equivalent to the `soft scale' in \cite{HT98,MY98,Yak98}), typically 
of order $30\,$GeV, compared to the central value $75\,$GeV 
chosen in previous works. This choice is motivated by the fact 
that the typical momentum transfer in the instantaneous interactions 
is of this order, or, if anything, smaller. 

The significant shift in the location of the peak impacts directly on 
the top quark mass measurement. There is no unique choice of the 
concept of the top quark mass. The top quark pole mass definition 
has been universally assumed in previous cross section calculations near 
the threshold; this is indeed an intuitively plausible choice as top 
quarks do not hadronize. However, the top quark pole mass definition 
is known to be more sensitive to non-perturbative effects \cite{BB94} 
than other mass definitions and the finite width of the top makes no 
difference in this respect. (The difference is, that the top quark pole 
mass is irrelevant, because a top quark is always off-shell by an 
amount $\sqrt{m_t\Gamma_t}$.). In \cite{Ben98} we argued that the shift 
of the peak position is related to large perturbative corrections which 
appear only, when the cross section is expressed in terms of the pole 
mass, and which have their origin in the sensitivity to distances larger 
than the toponium Bohr radius. 
The point is that the Coulomb potential in coordinate space and the 
pole mass receive the same large corrections \cite{Ben98,HSSW98}. We 
therefore perform a subtraction on the potential such that the 
subtracted terms are absorbed into a mass redefinition and at the 
same time cancel the large corrections to the pole mass. 
This leads to the `potential subtraction (PS) scheme' 
and the corresponding mass definition. For further details of the 
argument we refer to \cite{Ben98}. The PS mass at the subtraction scale 
$\mu_f$ is defined by 
\begin{equation}
\label{psms}
m_{t,\rm PS}(\mu_f) = m_t-\delta m_t(\mu_f),
\end{equation}
where
\begin{eqnarray}
\label{masssub}
\delta m_t(\mu_f) &=& -\frac{1}{2}\int\limits_{|\vec{q}\,|<\mu_f} 
\!\!\!\frac{d^3\bff{q}}{(2\pi)^3}\,[\tilde{V}(q)]_{\rm Coulomb} 
= \Delta(\mu_f)\Bigg[
1+\frac{\alpha_s}{4\pi}\left(a_1-b_0\left(\ln\frac{\mu_f^2}{\mu^2} 
-2\right)\right)
\nonumber\\
&&\hspace*{-1.5cm}+\,\frac{\alpha_s^2}{(4\pi)^2} 
\Bigg(a_2-\left\{2 a_1 b_0+b_1\right\}\left(\ln\frac{\mu_f^2}{\mu^2} 
-2\right)
+b_0^2\left(\ln^2\frac{\mu_f^2}{\mu^2}-4 \ln\frac{\mu_f^2}{\mu^2}+8
\right)\Bigg)\Bigg],
\end{eqnarray}
and $\Delta(\mu_f) = C_F\alpha_s \mu_f/\pi$. In the following, 
we re-express the $t\bar{t}$ cross section in terms of the PS mass. 
We suppose that $\mu_f$ scales as $m_t\alpha_s$ and count 
$\Delta(\mu_f)$ as order $m_t v^2$. Then all terms are re-expanded 
and terms beyond NNLO are dropped (modulo the resummation near the  
bound state poles discussed above). Note that $\Delta(\mu_f)$ is 
not expanded when it occurs in the combination $\sqrt{s}-2 
(m_{t,\rm PS}(\mu_f)+ \Delta(\mu_f))$. It should also not dominate this 
combination and this is why we do not use the $\overline{\rm MS}$ mass 
directly, which would lead to $\Delta \sim m_t \alpha_s$.

The peak in the $t\bar{t}$ cross section profile is, roughly speaking, 
the remnant of the first bound state pole (the `1S pole'). 
To understand the effect of the mass redefinition on the cross section 
qualitatively, it is useful to compute the correction to the 1S pole in 
terms of both mass definitions. The dominant correction in the pole 
mass scheme is related to the running coupling in the LO Coulomb 
potential. So let us take 
\begin{equation}
\delta\tilde{V}_k(\bff{q}) = \frac{-4\pi C_F\alpha_s}{\bff{q}^2} 
\left(-\frac{b_0\alpha_s}{4\pi} \ln\frac{\bff{q}^2}{\mu^2}\right)^{\!k}
\end{equation}
to compute the energy level shift
\begin{eqnarray}
\label{shift}
\delta E_k &=& \int \!\frac{d^3\bff{q}}{(2\pi)^3}\frac{d^3\bff{p}}{(2\pi)^3} 
\Psi_{\rm 1S}^*(\bff{p}+\bff{q}/2)\,\delta\tilde{V}_k(\bff{q})\,
\Psi_{\rm 1S}(\bff{p}-\bff{q}/2)
\nonumber\\
&=& \int \!\frac{d^3\bff{q}}{(2\pi)^3} \left(1+\frac{\bff{q}^2}{
(m_t C_F\alpha_s)^2}\right)^{\!-2}\,\delta\tilde{V}_k(\bff{q}).
\end{eqnarray}
The mass of the 1S bound state, including only the leading order Coulomb 
interaction and the above perturbation, is given by
\begin{equation}
M_{\rm 1S} = 2 m_t+E_1^{\rm LO}+ \delta E_k 
= 2 m_{t,\rm PS}(\mu_f) + E_1^{\rm LO,PS}(\mu_f) + \left[
\delta E_k+2\delta m_t(\mu_f)\right].
\end{equation}
Comparing (\ref{shift}) with (\ref{masssub}), we see that the integrands 
in $\delta E_k+2\delta m_t(\mu_f)$ cancel each other for 
$|\bff{q}| < m_t C_F \alpha_s$ and $|\bff{q}| < \mu_f$. At the same 
time the integral (\ref{shift}) 
is dominated by the contribution from small $\bff{q}$ 
quickly as $k$ increases. The result of this exercise for $\mu=30\,$GeV 
(in which case $m_t C_F\alpha_s(\mu)$ is also about $30\,$GeV) and 
$\mu_f=20\,$GeV is shown in the following table:
\begin{table}[h]
\addtolength{\arraycolsep}{0.1cm}
\renewcommand{\arraystretch}{1.4}
$$
\begin{array}{c|c|c|c|c|c|c|c|c}
\hline\hline
k 
& 1 & 2 & 3 & 4 & 5 & 6 & 7 & 8 \\ 
\hline 
\delta E_k/\mbox{MeV}  
& -489  & -214  & -109  & -78  & -67 & -70 & -85 & -118 \\
\hline
(\delta E_k+2\delta m_t)/\mbox{MeV}  
& +97   & +0.8  & +3.5 & +0.1 & +0.3 & 0 & 0 & 0 \\
\hline\hline
\end{array}
\vspace*{-0.2cm}
$$
\end{table}

\noindent
This shows that the prediction for the mass of the 1S state is stable 
in terms of the PS mass, and we expect a qualitatively similar 
conclusion for the top quark cross section. (The same observation 
can also be used to determine the bottom quark $\overline{\rm MS}$ 
mass from the $\Upsilon$ resonances accurately \cite{BSS98II}.) 
To demonstrate that the 
PS mass can also be related more accurately to the $\overline{\rm 
MS}$ mass $\bar{m}_t=m^{\overline{\rm MS}}_t(m^{\overline{\rm MS}})$, 
we assume $\bar{m}_t=165\,$GeV and obtain, adding loop corrections 
consecutively,
\begin{eqnarray}
m_t &=& \left[165.0+7.6+1.6+0.6\,(\mbox{est.})\right]\,\mbox{GeV}
\\
m_{t,\rm PS}(20\,\mbox{GeV}) &=& 
\left[165.0+6.7+1.2+0.3\,(\mbox{est.})\right]\,\mbox{GeV}.
\end{eqnarray}
The 2-loop correction follows from \cite{GRA} and (\ref{masssub}) and 
the 3-loop correction is based on an estimate in the so-called 
`large-$\beta_0$-limit' \cite{BBB}. Hence, the present uncertainty in 
the relation of the PS mass to the $\overline{\rm MS}$ mass is 
about $300\,$MeV.

\vspace*{0.2cm}


{\bf 4. Discussion.}
The $t\bar{t}$ cross section expressed in terms of 
$m_{t,\rm PS}(\mu_f)$ with $\mu_f=20\,$GeV is shown in the lower panel 
of Fig.~\ref{fig1}. Since the horizontal scale $E$ is now defined as 
$\sqrt{s}-2 m_{t,\rm PS}(20\,\mbox{GeV})$, we observe an overall, 
but trivial shift, related to the fact that 
$m_t-m_{t,\rm PS}(20\,\mbox{GeV})=1.75\,$GeV for $\mu=30\,$GeV. 
The important change is that in the PS scheme 
the location of the peak moves little 
as we go from LO to NLO to NNLO. Furthermore, the scale dependence 
of the peak location under variations of $\mu$ between $15$ and 
$60\,$ GeV is reduced by a factor of 2. 
The transition from the pole 
to the PS scheme has little effect on the shape and overall 
normalization of the cross section as expected. In particular a significant 
uncertainty of about $\pm 20\%$ in the normalization remains,  
larger than at NLO. 

The strong enhancement of the peak for the small 
scale $\mu=15\,$GeV is a consequence of the fact that the perturbative 
corrections to the residue of the 1S pole (see (\ref{residues})) 
become uncontrollable at scales not much smaller than $15\,$GeV. 
We find that these large corrections are mainly associated with the 
logarithms that make the coupling run in the Coulomb potential. This 
could be interpreted either as an indication that higher order 
corrections are still important (at such low scales) or that the terms 
associated with $b_0$ should be treated exactly, because they are 
numerically (but not parametrically) large. (If the Schr\"odinger equation 
with the potential (\ref{delV}) is solved exactly by numerical methods, 
the scale dependence of the peak height is indeed 
smaller \cite{HTinprep}.) Inspecting the logarithms in the result for 
the cross section, choosing an energy-dependent scale 
$\mu=2\,(m_t\,(E^2+\Gamma_t^2)^{1/2})^{1/2}$ would be most appropriate. 
Although this choice is our preferred one, we refrained from using it for 
the comparison of the pole and PS scheme, 
since the peak positions are at 
different energies in the two schemes. 

The uncertainties due to other  
parameters turn out to be less than the uncertainty due to the 
variation of the scale $\mu$. The dependence on $\alpha_s(m_Z)$ is 
discussed below. Varying $\mu_h$ by a factor of two around 
$m_t$ changes the cross section by a few percent near the peak. The 
effect of summing logarithms to NLL is of the same order. We have also 
checked the effect of some NNLL logarithms on the calculation and 
find a variation of the order of $\pm 5\%$. The relatively small effect of 
renormalization group improvement is a consequence of the absence of 
leading logarithms. We also checked the effect of varying $\mu_f$ 
between 5 and $40\,$GeV. From a purely pragmatic point of view, 
values of $\mu_f$ around $40\,$GeV lead to the most stable result. 
However, since $\mu_f<m_t C_F\alpha_s(\mu)\approx 30\,$ GeV is required 
from a conceptual point of view, we have chosen $\mu_f=20\,$GeV as our 
preferred setting. (From the point of view of non-perturbative infrared  
cancellations, it would be sufficient to choose $\mu_f$ larger than 
the strong interaction scale $1\,$GeV. However, from (\ref{shift}) we 
see that the cancellation becomes effective -- and the perturbative 
correction universal -- as soon as the integrand is dominated by $|\bff{q}| 
< m_t C_F\alpha_s$. For this reason it is legitimate and advantageous 
to choose $\mu_f$ significantly larger than the strong interaction scale.) 

\begin{figure}[t]
   \vspace{-4.8cm}
   \epsfysize=18cm
   \epsfxsize=12cm
   \centerline{\epsffile{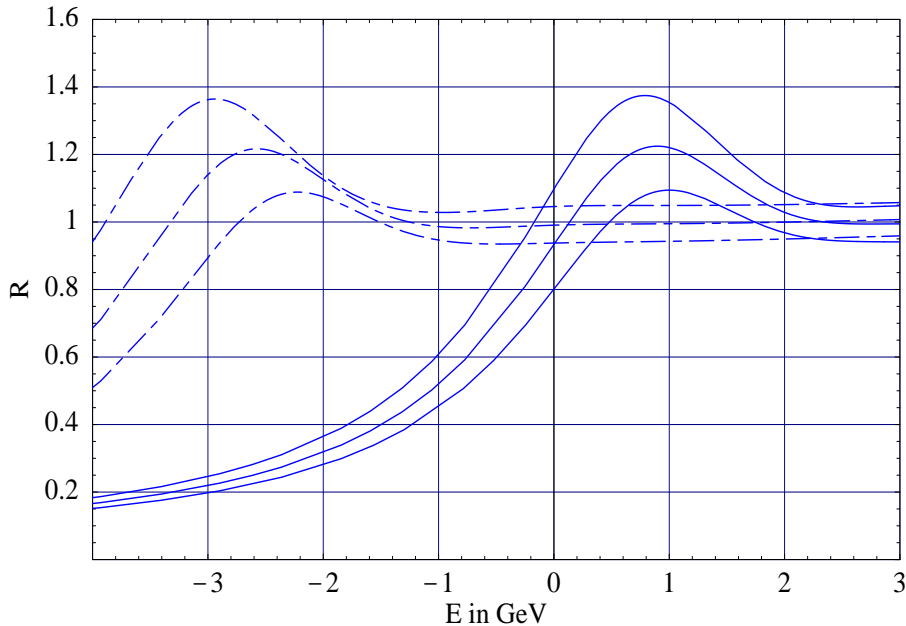}}
   \vspace*{-3.7cm}
\caption[dummy]{\label{fig2}\small 
Dependence of the NNLO $t\bar{t}$ cross section on $\alpha_s(m_Z)$ in the  
PS scheme (solid) and the pole scheme (long-short-dashed). The three 
curves in each scheme refer to $\alpha_s(m_Z)=0.113$ (lower), 
$\alpha_s(m_Z)=0.118$ (middle) and $\alpha_s(m_Z)=0.123$ (upper). 
Recall that $E=\sqrt{s}-2 m_t$ in the pole scheme but 
$E=\sqrt{s}-2 m_{t,\rm PS}(20\,\mbox{GeV})$ in the PS scheme. Other 
parameters: $m_t=\mu_h=175\,$GeV, $\Gamma_t=1.40\,$GeV, $\mu=30\,$GeV.}
\end{figure}

If we take (naively) the change in the peak position under scale 
variations as a measure of the uncertainty of the top mass measurement, 
we conclude that a determination of the PS mass with an error of 
about $100\,$-$150\,$MeV is possible. Given that the uncertainty in relating 
the PS mass to the $\overline{\rm MS}$ mass is about $300\,$MeV, 
this accuracy seems to be sufficient. 
We emphasize that it is not sufficient to invent an {\em ad hoc} mass 
definition in terms of which the peak position is stable empirically. 
In addition, such a 
mass definition needs to have a well-behaved relation order by order 
in perturbation theory to a mass definition 
relevant to top quark processes far away from threshold. For a 
realistic assessment of the error in the mass measurement, the theoretical 
line shape has to be folded with initial state radiation, beamstrahlung 
and beam energy spread effects. Since these effects are well understood, 
the main question that needs to be addressed is whether the normalization 
uncertainty leads to a degradation of the mass measurement after 
these sources of smearing are taken into account. This should be studied 
in a collider design specific setting.

In the PS scheme the correlation of the 
top quark mass with $\alpha_s$ is also strongly reduced, mainly because 
the perturbative corrections to the 1S pole are small in this 
scheme. This is indicated 
by Fig.~\ref{fig2}, which shows the dependence of the line shape 
on the value of $\alpha_s(m_Z)$. One can therefore rely less on 
input from top quark momentum distributions, which have been used in 
the pole scheme to constrain $m_t$ and $\alpha_s(m_Z)$ simultaneously. 
Since momentum distributions are more sensitive to non-perturbative 
effects than the inclusive cross section, this is another advantage 
of the PS scheme.

In conclusion, we evaluated the next-to-next-to-leading order QCD 
correction to top quark production near threshold in the conventional 
pole scheme and in the PS scheme \cite{Ben98}. We employed 
factorization in dimensional regularization and summed next-to-leading 
logarithms in the top quark velocity. We find that the cross section 
expressed in the PS scheme allows us to determine the top PS mass more 
accurately than the top pole mass. The PS mass, in turn, can also be 
related more accurately to the top quark $\overline{\rm MS}$ mass.

\vspace*{0.2cm}

{\bf Acknowledgements.} We thank A.H.~Hoang for extensive discussions and 
for comparing numerical results of their calculation \cite{HTinprep} prior 
to publication. The work of M.B. is supported in part by the 
EU Fourth Framework Programme `Training and Mobility of
Researchers', Network `Quantum Chromodynamics and the Deep Structure 
of Elementary Particles', contract FMRX-CT98-0194 (DG 12 - MIHT). 
The work of V.S. has been supported by the Volkswagen Foundation, 
contract No.~I/73611, the Russian Foundation for
Basic Research, project 98--02--16981 and by INTAS, grant 93--0744.

\end{document}